# Unraveling Quantum Size-Dependent Optoelectrical Phenomena in Hot Carrier Quantum Well Structures

Nil Selen Aydin[1], Leopold Rothmayer[1], Nabi Isaev[1], Pavel Avdienko[1], Nori N. Chavira Leal[2], Kai Müller[1,2], Jonathan J. Finley[1], Gregor Koblmüller[1,3], Hamidreza Esmaielpour[1]

[1] Walter Schottky Institut, TUM School of Natural Sciences, Technical University of Munich, 85748 Garching, Germany.

[2] TUM School of Computation, Information and Technology, Technical University of Munich, 80333 Munich, Germany.

[3] Institute of Physics and Astronomy, Technical University Berlin, 10623 Berlin, Germany.

**Abstract** – The enhancement of power conversion efficiency beyond the theoretical limit of single-junction solar cells is a key objective in the advancement of hot carrier solar cells. Recent findings indicate that quantum wells (QWs) can effectively generate hot carriers by confining charged carriers within their potential wells and by optimizing material properties. Here, we investigate the impact of quantum confinement on the thermodynamic properties of photogenerated hot carriers in p-i-n InGaAs/InAlAs heterostructure diodes, utilizing QW thicknesses of 4 nm, 5.5 nm, and 7.5 nm. The optical properties of these nanostructures reveal significant hot carrier effects at various lattice temperatures, with a pronounced effect noted at lower temperatures. The experimental results indicate that the widest QW exhibits stronger hot carrier effects than the thinner QWs. Additionally, the open-circuit voltage of the samples demonstrates a correlation with the degree of quantum confinement, mirroring trends observed in the quasi-Fermi level splitting of hot carriers. However, the magnitudes recorded exceed the bandgap of the quantum structures, suggesting that this behavior may be influenced by the barrier layer. Furthermore, the short-circuit current of the samples reveals a strong dependence on excitation power, but not on the degree of quantum confinement. This indicates that the majority of the photocurrent is generated in the barrier, with negligible contributions from photogenerated carriers within the QWs. This study provides insights into the role of quantum confinement on the opto-electrical properties of non-equilibrium hot carrier populations in QW structures.

## I. INTRODUCTION

Hot carrier solar cells represent a promising avenue for surpassing the theoretical efficiency limit of 33% established for single-junction photovoltaic (PV) devices [1]. A significant challenge to achieving higher efficiency is the thermalization of hot carriers within semiconductors. Hot carriers, which are high-energy particles produced by the absorption of photons with energies exceeding the bandgap of the solar cell absorbers, undergo rapid thermalization [2,3]. This process presents difficulties in harnessing their excess kinetic energy for electricity generation.

Extracting hot carriers from quantum well (QW) structures has emerged as an effective approach to address the Shockley-Queisser efficiency limit [4,5,6]. These quantum structures are capable of decelerating the rates at which hot carriers relax, thereby creating favorable conditions for charge transport through energy-selective contacts. Several key mechanisms play a role in mitigating the rapid thermalization of hot carriers within these nanostructures. These mechanisms include the spatial confinement of hot carriers in potential wells [7,8], the formation of discrete energy levels due to quantum confinement [9,10], and the phonon-bottleneck effect achieved through material property optimization [11,12]. QW structures from different material systems, such as GaAs/AlGaAs QWs [13,14], InAs/AlAsSb QWs [15], InGaAs/InAlAs QWs [16,17], InGaAsP single QWs [4,18], and InGaN/GaN QWs [19], have demonstrated promising potentials in slowing the rates of hot carrier thermalization.

In addition to reducing thermalization rates, the development of efficient energy-selective contacts is crucial for the effective operation of hot carrier solar cells [1,20]. Such structures are designed to selectively block the passage of low-energy carriers, commonly referred to as cold carriers, while allowing hot carriers to be transmitted. Strategies to realize this capability may include the use of wide-bandgap semiconductors [21,22], resonant-tunneling double barriers [23], superlattice quantum wells [24], and the incorporation of high-energy satellite valleys [25,26].

Despite the encouraging findings, several challenges remain, including competing thermalization mechanisms in the QW structures and the intricate transport dynamics of hot carriers within these nanostructures [5,11,27]. Additionally, the presence of discrete energy levels in QWs and band-filling effects complicates the accurate determination of the temperature of hot carriers [16,28]. As a result, the application of advanced analytical techniques, such as full-spectral fitting, is of paramount importance for investigating the thermodynamic properties of hot carriers through optical spectroscopy in these nanostructures.

This study focuses on examining the strength of QW size effects on the properties of hot carriers in epitaxially grown InGaAs single-QW structures by varying their thicknesses. Hereby, optoelectrical characterizations are conducted to investigate the properties of hot carriers within the nanostructures, with particular emphasis on the correlation between the thermodynamic properties of these non-equilibrium populations and the current extracted from the QWs.

## II. EXPERIMENTAL METHODS AND DISCUSSIONS

This study explores hot carrier effects in QW structures consisting of $In_{0.53}Ga_{0.47}As$ wells embedded within $In_{0.52}Al_{0.48}As$ barriers. These structures were grown lattice-matched to a p$^+$-doped ($4 \times 10^{18} cm^{-3}$) InP(001) substrate using molecular beam epitaxy, as shown in Figure 1(a). We fabricated three distinct samples, each with a different QW layer width: 4 nm, 5.5 nm, and 7.5 nm. Each structure includes a 100 nm $In_{0.52}Al_{0.48}As$ barrier on both sides of the well, ensuring that all samples have identical confining environments. Additionally, the QW structures are capped with a 100 nm n$^+$-doped ($1 \times 10^{19} cm^{-3}$) $In_{0.52}Al_{0.48}As$ layer, which serves as an electron contact. This design facilitates the collection of photo-generated hot carriers from the system. By varying only the QW width in these structures, we enable a direct comparison of hot carrier dynamics as a function of size confinement effects. The band-energy diagram of the single-QW structures is illustrated in Figure 1(b).

To conduct a comprehensive investigation into the optoelectronic properties of hot carriers in this set of samples, photovoltaic test devices were realized by post-growth fabrication. As depicted in Figure 1(a), metal contacts comprised of Ti/Pt/Au (10 nm/50 nm/100 nm) were deposited on both n-doped and p-doped regions, followed by an annealing process conducted at 150 °C for 90 seconds. Using a phosphoric acid solution (H$_3$PO$_4$:H$_2$O$_2$:H$_2$O / 1 ml:1 ml:38 ml), mesa structures that result in diodes with dimensions of 250 μm x 350 μm were realized, featuring a 100 μm opening designed for laser excitation.

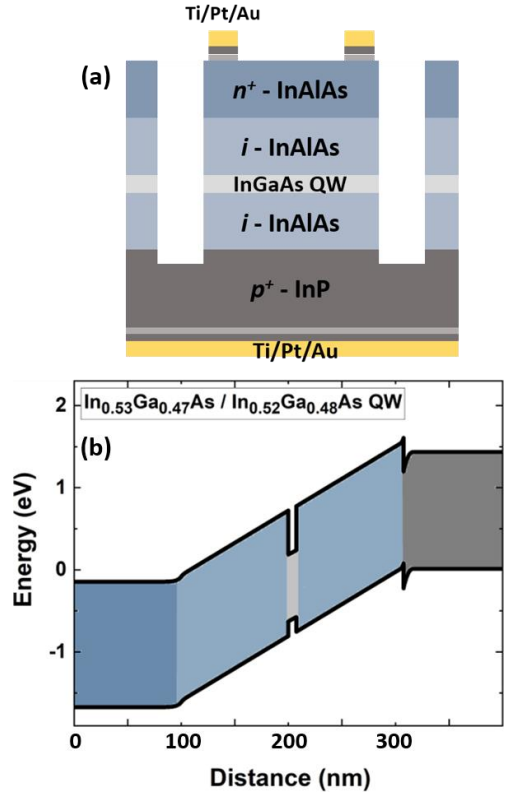

Figure 1. (a) Schematic of the p-i-n InGaAs QW mesa structures with metal contacts for electrical studies. (b) Band-energy diagram of the entire QW heterostructure.

The investigation into the hot carrier properties of the QW structures is performed utilizing micro-photoluminescence (μ-PL) spectroscopy at varying excitation power levels and lattice temperatures. The optical configuration comprises a Ti-Sapphire laser operating at a 740 nm excitation wavelength with a 2 μm spot size, a TRIAX550 spectrometer, and a liquid-nitrogen-cooled Horiba IGA-3000V InGaAs detector for capturing the PL emission from the samples. The QW structures are situated within a cryostat that can achieve temperatures as low as 4.2 K through a liquid helium cooling system. This arrangement facilitates precise control over the temperature of the samples, allowing for comprehensive measurements of the photogenerated hot carrier properties at various lattice temperatures.

Figure 2 presents the PL spectra emitted by the QWs of varying thicknesses: (a) 4 nm, (b) 5.5 nm, and (c) 7.5 nm, measured at 10 K under different excitation powers. The data indicates a redshift in the PL emission correlated with an increase in the thickness of the QWs, i.e. a peak emission shift from ~1.03 eV (4 nm QW) to ~ 0.91 eV (7.5 nm QW) at the lowest excitation power. This phenomenon is attributed to the reduction in the confined bandgap energy resulting from the widening of the potential wells. Furthermore, it is noted that as the excitation power increases, the slope on the high-energy side of the spectra becomes less steep, indicating the generation of hot carriers within the system. To achieve a quantitative assessment of the thermodynamic properties of the hot carriers in the quantum structures, the PL spectra are fitted utilizing the generalized Planck's radiation law [29]:

$$I_{PL}(E) = \frac{2\pi A(E) (E)^2}{h^3 c^2} \left[ exp\left(\frac{E - \Delta\mu}{k_B T}\right) - 1 \right]^{-1}, \tag{1}$$

where, "$I_{PL}$" is the PL intensity, "$A(E)$" the energy-dependent absorptivity, "$h$" the Planck constant, "$k_B$" the Boltzmann constant, "$c$" the speed of light. The temperature and the quasi-Fermi level splitting of hot carriers are represented by "$T$" and "$\Delta\mu$", respectively, in this equation. By conducting a comprehensive analysis of the PL spectrum, one can ascertain both the absorptivity of the sample and the characteristics of photo-generated hot carriers in the quantum structures [28]. This fitting methodology emphasizes the significance of the band-filling effect within the absorptivity term, particularly at high-excitation powers. Additionally, the application of a full-spectrum fit is particularly critical for nanostructured materials, given that their absorption spectra exhibit energy dependence that extends beyond the bandgap [16].

Performing full-spectral fitting requires modeling the absorptivity of the QW structures, which includes the discrete energy levels created by quantum confinement effects. The absorptivity, A(E), of the samples is described by:

$$A(E) = [1 - R(E)] \cdot (1 - exp[-(\alpha_w d_w + \alpha_b d_b)]), \tag{2}$$

where "$R$" is reflectivity, "$\alpha$" is the absorption coefficient, and "$d$" is the absorber thickness. The indices "$w$" and "$b$" refer to the quantum well and the barrier, respectively. The absorption coefficient "$\alpha_{w0}$" of the QW is given by [30,31]:

$$\alpha_{w0}(E) = a_x \cdot exp\left[-\frac{(E - E_x)^2}{2\Gamma_x^2}\right] + \sum_{i=1}^{n} a_i \cdot \frac{1}{1 + exp\left(-\frac{E - E_i}{\Gamma_i}\right)} \cdot \frac{2}{1 + exp\left(-2\pi\sqrt{\frac{R_y}{|E - E_i|}}\right)}, \tag{3}$$

and the absorption coefficient "$\alpha_{b0}$" of the barrier:

$$\alpha_{b0}(E) = a_b \cdot \frac{1}{1 + exp\left(-\frac{E - E_b}{\Gamma_b}\right)}. \tag{4}$$

where "$a$" is the amplitude, "$E$" is the optical transition energy, "$\Gamma$" the spectral linewidth broadening, and "$R_y$" the effective Rydberg energy. The first and second terms in equation (3) correspond to the excitonic transitions and the band-to-band transitions, respectively, originating from multiple discrete energy levels within the QW structures.

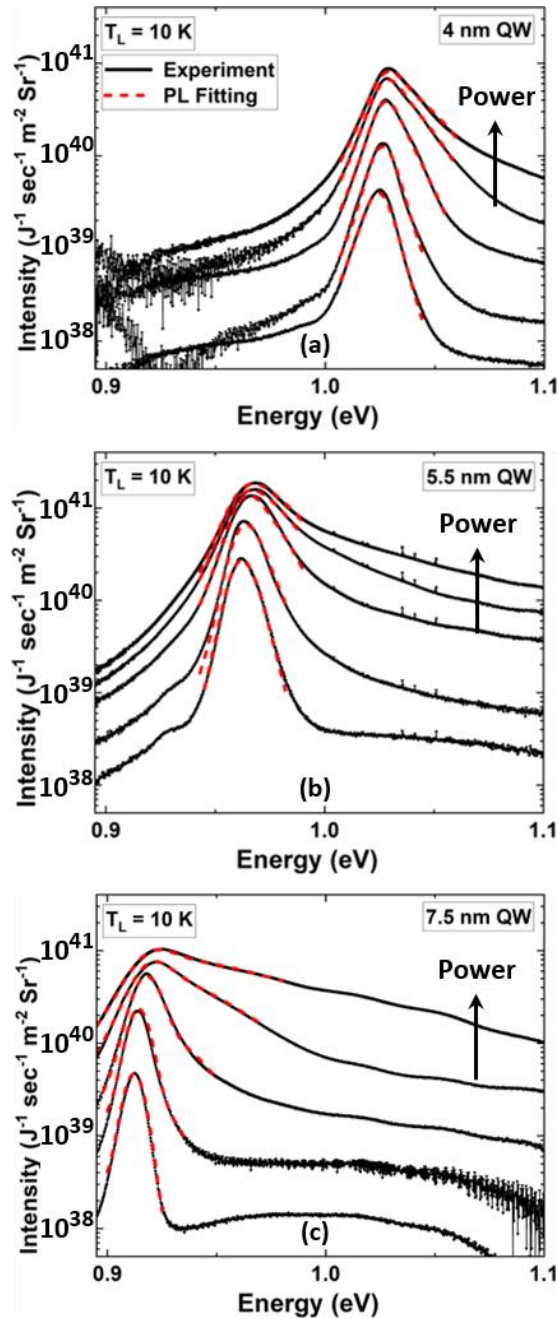

Figure 2. Excitation power-dependent PL spectra at 10 K of (a) 4 nm, (b) 5.5 nm, and (c) 7.5 nm thin QWs. The black solid lines and the red dashed lines illustrate the experimental and the full PL fitting results, respectively. All measurements were performed at 10K.

Upon photo-excitation, the availability of states in both the conduction and valence bands diminishes, resulting in a corresponding decrease in absorptivity, which can significantly influence the optical properties of materials. This phenomenon arises from the filling of energy levels near the band edge by an increased density of photo-generated carriers. In accordance with the Pauli exclusion principle, the remaining available states shift to higher energy levels. As a result, the absorptivity of the sample becomes contingent upon the thermodynamic properties, including temperature and the quasi-Fermi level splitting, which can be described by [4]:

$$\frac{\alpha}{\alpha_0} = f_v^e - f_c^e = \frac{\sinh\left(\frac{E - \Delta\mu}{2k_BT}\right)}{\cosh\left(\frac{E - \Delta\mu}{2k_BT}\right) + \cosh\left[\frac{m_h - m_e}{m_h + m_e} \cdot \frac{E - E_g}{2k_BT} - \frac{1}{2}\ln\left(\frac{m_h}{m_e}\right)\right]}. \quad (5)$$

where "$f_v^e$" and "$f_c^e$" indicate the Fermi-Dirac distribution of carriers in the conduction and the valence bands, respectively. The effective mass of holes and electrons in the system are represented by "$m_h$" and "$m_e$". The results of full PL spectral fitting for the QW emission are shown by red-dashed lines in Figure 2.

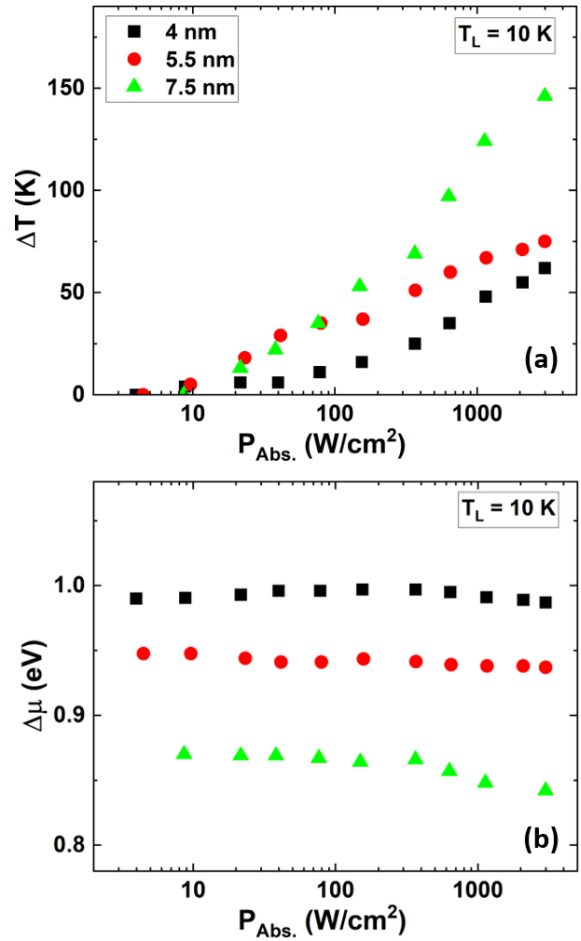

Figure 3. (a) The temperature and (b) the quasi-Fermi level splitting of the QW structures of various thicknesses versus the absorbed power density at 10 K.

The results of the hot carrier temperature (ΔT: the temperature difference between the hot carriers and the lattice temperature) and the quasi-Fermi level splitting as a function of excitation power at 10 K are presented in Figures 3 (a) and (b) for the different QW structures. It is observed that an increase in absorbed power density correlates with a rise in hot carrier temperature, suggesting a more pronounced hot carrier effect at elevated excitation powers. Notably, the temperature of the hot carriers in the thickest QW (7.5 nm) exceeds that in the other samples, see Figure 3(a). The observed increase in hot carrier effects associated with QW thickness is consistent with the theoretical predictions established for analogous material systems. This effect is primarily attributed to the interplay between intra-subband and intersubband transitions, as well as the influence of electrostatic screening within the QW structures. Notably, these effects exhibit a dependence on the QW thickness and demonstrate an increase in hot carrier temperature versus the size of the QWs, particularly within the sub-10 nm range [10,32].

Another mechanism that can influence the relaxation rates of hot carriers in quantum wells is based on interface states, which can accelerate the thermalization rates of these carriers, ultimately leading to reduced hot carrier effects within the system. [8,33,34] Evidence of the influence of interface states is highlighted by the linewidth broadening observed in the PL spectrum, particularly on the low-energy side. Figure S1 in the Supplementary Materials presents an analysis of the relationship between the PL linewidth broadening and the thickness of the QWs. It is observed that as the thickness of the QW increases, the linewidth broadening exhibits a decline. This trend can be attributed to the influence of interface roughness between the QW and the barrier, which becomes less significant in its effect on the PL spectrum by increasing the QW thickness [35,36]. Notably, this effect scales with the inverse of the square of the thickness ($1/L_z^2$) [35], as indicated in the inset of Figure S1(b).

The results of the quasi-Fermi level splitting in the QW structures shown in Figure 3(b) indicate that the magnitude of the splitting increases at a given absorbed power density when the thickness of the QWs is reduced. This phenomenon is due to the reduction in the confined energy states of the samples, which occurs as the potential wells widen.

The electrical properties of the QW structures were investigated by the micro-PL setup across a range of lattice temperatures and excitation powers. Figure 4(a) presents the current-voltage (J-V) characteristics of a 7.5 nm QW at 10 K under various excitation power levels. The conditions for the optical and electrical measurements were kept consistent, facilitating a meaningful correlation between the thermodynamic properties of photo-generated hot carriers and the electrical results observed. As indicated in Figure 4(a), an increase in excitation power density leads to a corresponding rise in photocurrent, resulting in a higher short-circuit current ($J_{SC}$). Furthermore, an increase in the open-circuit voltage ($V_{OC}$) with power is noted, which aligns with the electrical characteristics typically associated with diodes in this system [37].

The dark J-V characteristics of the 7.5 nm QW p-i-n diode at various lattice temperatures are illustrated in Figure 4(b). The data indicates that an increase in lattice temperature results in a red shift of the J-V characteristics, which correlates with the decrease in bandgap energy as temperature rises [38]. Furthermore, the semi-logarithmic representation of the J-V curve demonstrates that there is minimal leakage current under reverse bias. This observation is important for accurately assessing the contribution of photocurrent to the short-circuit current. To gain deeper insights into the electrical properties of the QW diode, the ideality factor is determined using the standard diode equation, as given by [37]:

$$J(V) = J_s \, exp\left(\frac{qV}{nk_BT} - 1\right), \tag{6}$$

where the term "$J_s$" represents the reverse saturation current density. Ideality factors ranging from 1 to 2 indicate that the dominant current in the system is due to diffusion and recombination processes.

However, if the ideality factor is greater than > 2, it suggests that a tunneling mechanism is prevailing in the diode current of the system [39].

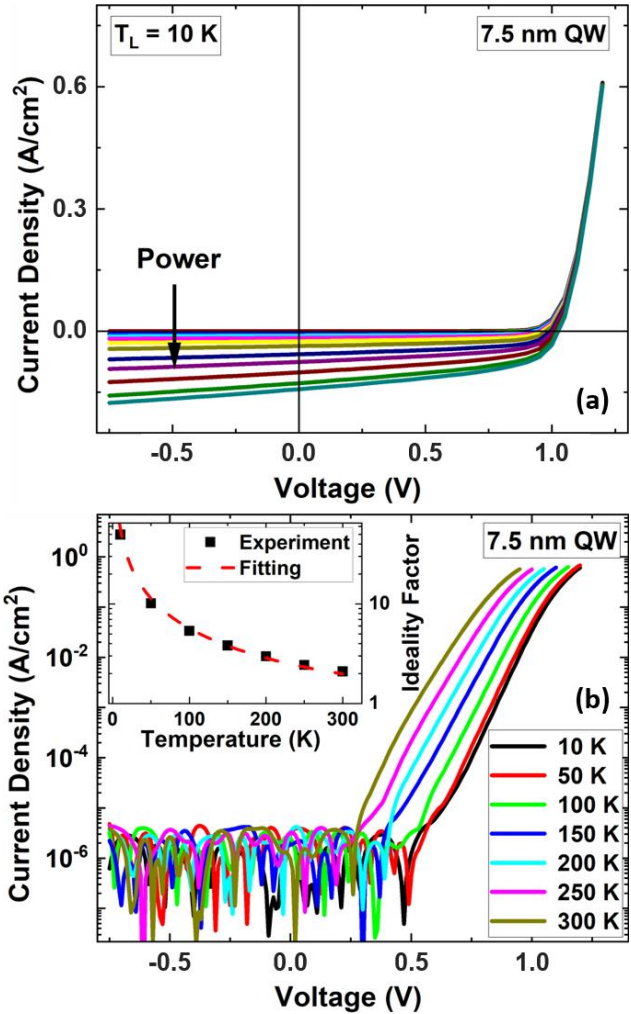

Figure 4. (a) The current-voltage characteristics of the 7.5 nm QW at 10 K under different excitation power levels. (b) Current-voltage curves of the QW in the dark at various lattice temperatures. The inset shows the relationship between the ideality factor and lattice temperature, as determined by the standard ideal diode equation. The red dashed line represents the fitting results compared to the experimental data.

The inset of Figure 4(b) illustrates the relationship between the ideality factor and lattice temperature. At room temperature, the ideality factor is approximately 2, suggesting that the recombination mechanism in the active region has a significant impact on charge transport [37]. Notably, as the lattice temperature decreases, the ideality factor exhibits a marked increase, reaching around 50 at 10 K. This behavior may stem from alterations in charge transport due to the recombination of charge carriers within trap states at the interface [39]. To further investigate the origins of this unusually high ideality factor, experimental results are compared with a tunneling-enhanced interface recombination model, as described by [40]:

$$n = \frac{E_{00}}{k_B T} \coth\left(\frac{E_{00}}{k_B T}\right), \tag{7}$$

where "$E_{00}$" is the characteristic tunneling energy. The results demonstrate a strong correlation between the experimental findings and the theoretical model, with the characteristic tunneling energy estimated at 49 meV. This observation indicates that, at lower temperatures, carrier recombination in localized states at the heterostructure interfaces is the primary mechanism at play, leading to significant barriers for charge transport within the system. As the lattice temperature increases, carriers gain thermal energy, enabling them to escape from these localized states and migrate toward the contact layers [40]. These trap states may be induced by non-idealities during the sample growth process, potentially diminishing charge collection efficiency at reduced temperatures.

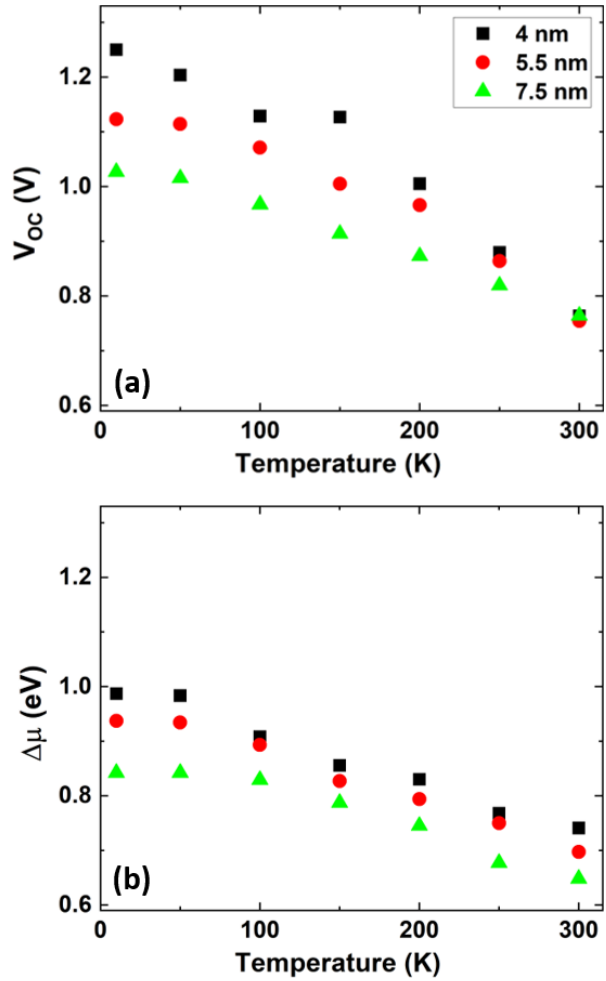

Figure 5. Temperature-dependent open-circuit voltage (a) and the quasi-Fermi level splitting (b) for the QW structures of various thicknesses under 3 $kW/cm^2$ absorbed power density.

The results of the open-circuit voltage for the QW structures versus the lattice temperature at a specified absorbed power density (3 $kW/cm^2$) are illustrated in Figure 5(a). It is observed that as the lattice temperature increases, the magnitude of the voltage shifts to lower values. This observation aligns with the red-shift behavior of the bandgap energy that occurs at elevated temperatures. Furthermore, the open-circuit voltage for the thinnest QW (4 nm) exhibits higher values compared to thicker QWs. While the open-circuit voltage does depend on the thickness of the QWs, its magnitude remains greater than the bandgap of the QW structures, as shown in Figure 2.

To compare the electrical behavior of the diodes with the thermodynamic properties of hot carriers, the results of the quasi-Fermi level splitting for the QW structures under the same absorbed power density are presented in Figure 5(b). The quasi-Fermi level splitting of hot carriers demonstrates similar behavior to the open-circuit voltage in relation to lattice temperature. Despite the similarities in the temperature dependence of both the open-circuit voltage and the quasi-Fermi level splitting of the QWs, the larger values of the open-circuit voltage than the QW bandgap ($V_{oc} > E_g/q$), may indicate that the $V_{oc}$ is more closely related to the quasi-Fermi level splitting of the barrier rather than the QWs themselves.

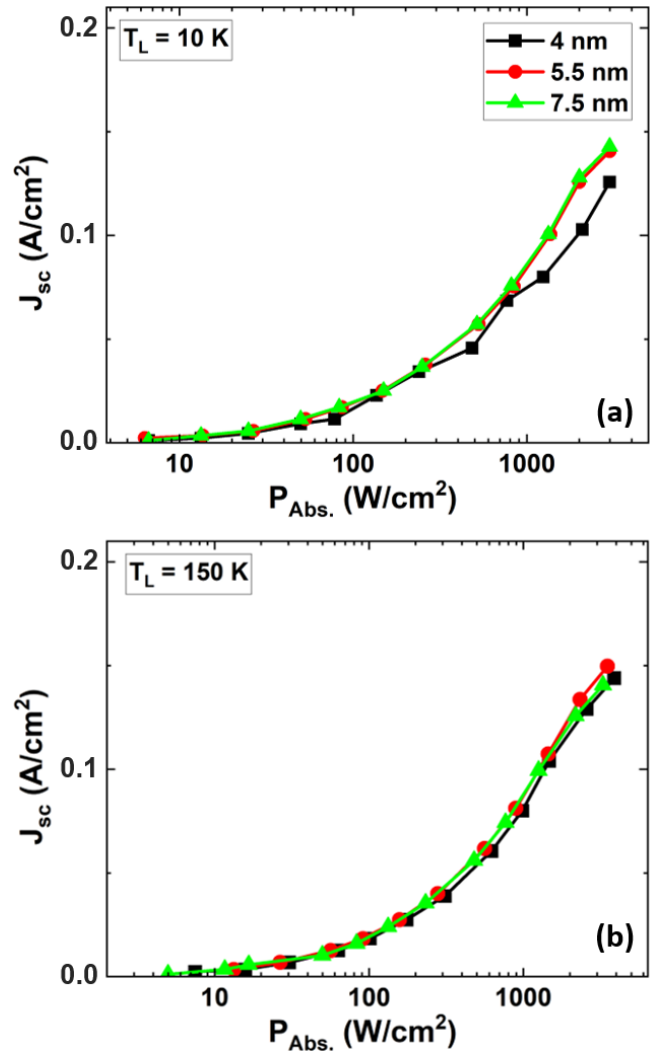

Figure 6. The results of short-circuit current density versus the absorbed power density of the QW structures at (a) 10 K, and (b) 150 K.

To further investigate the optoelectronic properties of the QW structures and their relationship with photo-generated hot carriers, we plot the short-circuit current of the QW diodes at 10 K and 150 K, as shown in Figures 6 (a) and (b), respectively. The results indicate that as the absorbed power density increases, the magnitude of the short-circuit current density also increases. However, the behavior observed is notably similar across various QW structures at both lattice temperatures. It is anticipated that if the photocurrent is primarily due to hot carrier extraction, there should be a correlation between the hot carrier temperature and the quasi-Fermi level splitting with the short-circuit current density. However, no significant dependence of the short-circuit current on the size of the quantum confinement is observed in the samples.

To study the origin of this effect, we determine the amount of photo-absorption of the QW and the barrier layers. The simulation results obtained from Lumerical software® indicate that the level of photo-absorption at an excitation wavelength of 740 nm is approximately 60 times greater in the barrier than in the QW region, as shown in Figure S2 in the Supplementary Materials. Consequently, it can be inferred that the photo-generated hot carriers within the QWs contribute negligibly to the photocurrent of the diodes. Additionally, the optical emission from the InAlAs barrier layers shows photoluminescence around 750 nm, as detailed in the supplementary information (see Figure S3). As the QW structures are excited using a 740 nm laser, it is anticipated that the photo-generated carriers in the barrier layers will have a dominant role in contributing to the photocurrent of the samples.

Future studies aimed at enhancing the effects of hot carriers will focus, thus, on several key approaches. One effective method involves increasing light absorption in the QWs by implementing pumping strategies below the barrier. Additionally, positioning the potential well closer to the surface may result in a higher density of hot carriers, thus leading to more robust thermionic emission for their extraction. This, in turn, can significantly improve the contribution of hot carriers to the photocurrent of such nanostructures. Furthermore, the design of energy-selective contacts, particularly for the extraction of hot electrons, is advantageous for a comprehensive study of the electrical properties associated with hot carriers within the system.

### III. CONCLUSION

In summary, this study examined the influence of the quantum well size on hot carrier properties in InGaAs/InAlAs QW heterostructures through a series of optoelectrical characterizations. The photoluminescence spectra of the QW structures were analyzed using a full-spectral fitting approach based on the generalized Planck's radiation law. This method enabled us to determine the temperature and the quasi-Fermi level splitting of the photogenerated hot carriers within the system. Our findings indicate that, upon photoexcitation, evidence of hot carrier distributions was observed. The QW structure with a thickness of 7.5 nm exhibited the highest hot carrier temperature, aligning with theoretical predictions for similar heterostructure designs. Additionally, interface roughness was observed in the QWs, which correlates with the thickness of the potential well. Consequently, reducing the width of the quantum well leads to increased rates of hot carrier thermalization, resulting in weaker hot carrier effects for thinner QWs.

The current-voltage characteristics of the QW heterostructures were then systematically investigated under excitation powers and lattice temperatures that are consistent with those used in optical spectroscopy. This methodology facilitates the correlation of thermodynamic properties of photo-generated hot carriers with electrical signals. The analysis of open-circuit voltage reveals a decrease in its magnitude as the lattice temperature increases, which is consistent with observations related to quasi-Fermi level splitting. Furthermore, the QW structure with a thickness of 4 nm exhibited higher voltage values compared to the thicker QW structures, supporting the implications of quantum confinement effects within the system. It is noteworthy that the magnitudes of the open-circuit voltage exceed the bandgap of the samples ($V_{oc} > E_g/q$), indicating that the open-circuit voltage is predominantly influenced by the quasi-Fermi level splitting of the barrier rather than that of the QWs themselves. Additionally, the short-circuit current of the QW structures demonstrates a clear dependence on the excitation power, with the short-circuit current ($J_{sc}$) increasing with the excitation power. However, no significant dependence on the QW thickness is observed, as all QW diodes display similar current magnitudes in relation to power variations. This phenomenon can be explained by the larger number of photogenerated carriers produced in the barrier region, which exhibits greater photoabsorption compared to the QWs. As a result, the barrier makes a more significant contribution to the short-circuit current in the system.

## ACKNOWLEDGEMENT

This work was supported by the Deutsche Forschungsgemeinschaft (German Research Foundation (DFG)) under Germany's Excellence Strategy via the Cluster of Excellence e-conversion (EXC 2089/1-390776260)). The authors further acknowledge support from Marie Sklodowska Curie Action (MSCA) via the TUM EuroTechPostdoc2 Grant Agreement (No. 899987). The authors further thank H. Riedl for support with the MBE system.

# Unraveling Quantum Size-Dependent Optoelectrical Phenomena in Hot Carrier Quantum Well Structures

Nil Selen Aydin[1], Leopold Rothmayer[1], Nabi Isaev[1], Pavel Avdienko[1], Nori N. Chavira Leal[2], Manuel Rieger[1], Kai Müller[1,2], Jonathan J. Finley[1], Gregor Koblmüller[1,3], Hamidreza Esmaielpour[1]

[1] Walter Schottky Institut, TUM School of Natural Sciences, Technical University of Munich, 85748 Garching, Germany.

[2] TUM School of Computation, Information and Technology, Technical University of Munich, 80333 Munich, Germany.

[3] Institute of Physics and Astronomy, Technical University Berlin, 10623 Berlin, Germany.

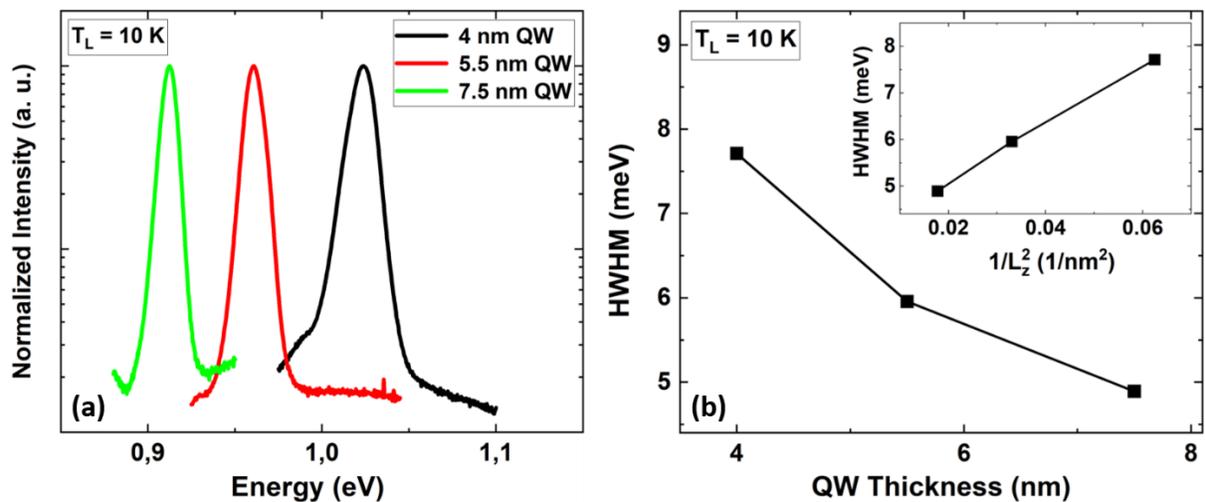

Figure S1. (a) The normalized photoluminescence spectra of the quantum wells with thicknesses of 4 nm (black), 5.5 nm (red), and 7.5 nm (green) at a temperature of 10 K under low excitation power. (b) The half-width at half-maximum (HWHM) measured from the low-energy side of the PL spectra versus the QW thickness. It is observed that increasing the thickness of the QWs leads to a reduction in linewidth broadening. The inset illustrates the dependence of the HWHM versus the inverse of the QW thickness square ($1/L_z^2$). The results indicate a linear dependence between the HWHM and $1/L_z^2$, which reflects how the exciton localization energy changes in relation to the roughness of the interface.

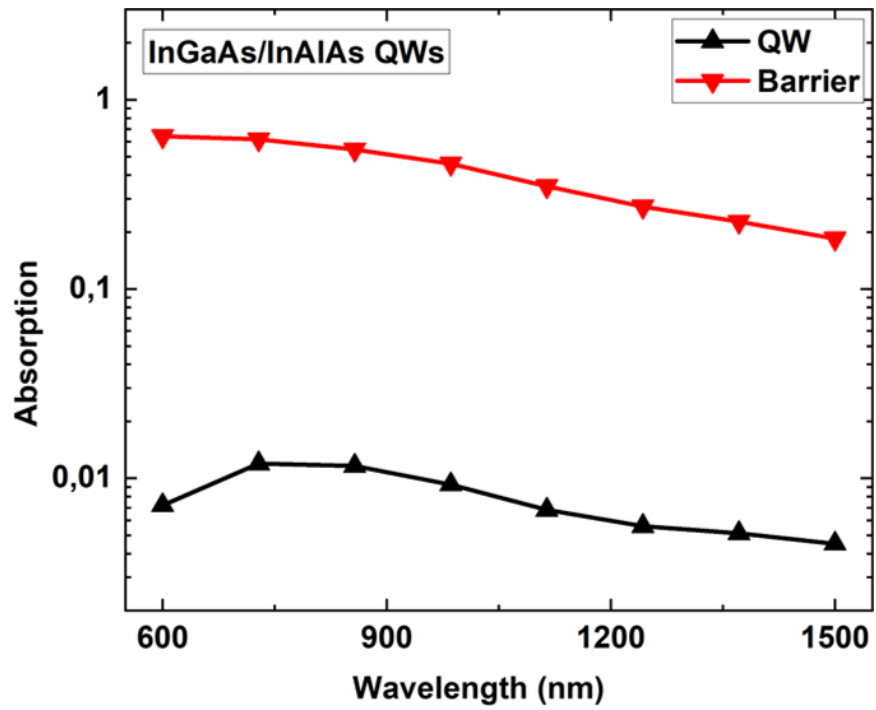

Figure S2. The absorption characteristics of the 7.5 nm QW heterostructure in the potential well (black) and in the barrier (red) regions. The absorption of the other samples is almost identical.

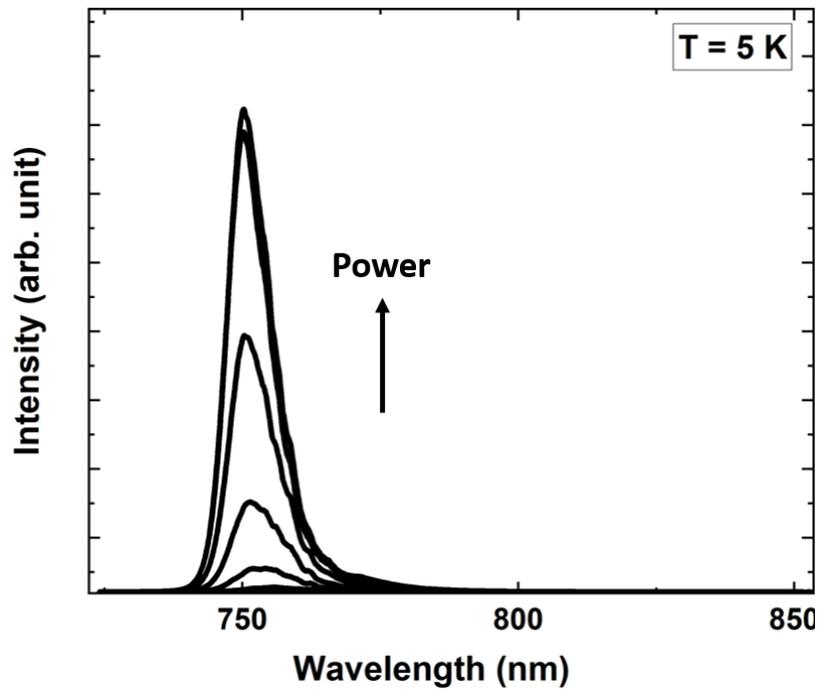

Figure S3. The PL spectra of the InAlAs barrier emitted at 5 K under various excitation powers.